\documentclass[12pt]{article}
\usepackage{a4}
\usepackage{amsfonts,amssymb,amsmath,amsthm}
\usepackage{graphicx}

\newcommand{\Dir}{{\not{\hspace{-4pt} D}}}

\begin{document}
\title{Heat kernel, effective action and anomalies
in noncommutative theories}

\author{ Dmitri V. Vassilevich\thanks{Also at V.~A.~Fock Institute of
Physics, St.Petersburg University, Russia; email:
Dmitri.Vassilevich@itp.uni-leipzig.de}\\
{\it Institut f\"{u}r Theoretische Physik, Universit\"{a}t
Leipzig,}\\
{\it D-04109 Leipzig, Germany } }

\maketitle
\begin{abstract}
Being motivated by physical applications (as the $\phi^4$ model)
we calculate the heat kernel coefficients for generalised Laplacians
on the Moyal plane containing both left and right multiplications.
We found both star-local and star-nonlocal terms. By using these results
we calculate the large mass and strong noncommutativity expansion
of the effective action and of the vacuum energy. We also study the
axial anomaly in the models with gauge fields acting on fermions from
the left and from the right. 
\end{abstract}
\section{Introduction}
Noncommutative field theories attract an ever increasing attention
of the researches (see the reviews \cite{DoNe,Szabo,Schaposnik:2004rc}).
Various forms of noncommutativity arise from strings, gravity, deformation
quantisation, and quantum Hall effect. Therefore, the
possibility that our space-time
is a noncommutative one should be taken seriously.

The heat kernel method \cite{Vassilevich:2003xt,NewGilkey,KirstenBook}
is one of the powerful instruments of modern quantum field theory.
One considers an operator $D$ which defines the spectrum of quantum
fluctuations on a given background. For a bosonic theory, $D$ is typically
a second order operator of Laplace type. One defines the heat kernel
(or, more precisely, the heat trace) for $D$ as an $L_2$ trace
\begin{equation}
K(t,D)={\mathrm{Tr}}_{L^2} \left( \exp (-tD) \right) \,,\label{heattr}
\end{equation}
where $t$ is a spectral parameter. Usually, (see 
\cite{Vassilevich:2003xt,NewGilkey,KirstenBook} for more precise statements)
the heat trace is well defined for $t>0$, and, as $t\to +0$, there is
a full asymptotic expansion  
\begin{equation}
K(t,D)\simeq \sum_{k=0}^\infty t^{(k-n)/2} a_k (D). \label{asymptotex}
\end{equation}
On manifolds without boundaries the odd-numbered coefficients vanish,
$a_{2p+1}=0$, and the even-numbered coefficients can be expressed 
via integrals over the manifold of local invariants constructed from the
symbol of $D$.

In quantum field theory the coefficients in the expansion (\ref{asymptotex})
(called also the heat kernel coefficients) define the one-loop
divergences, anomalies, and various expansions of the effective action.
In noncommutative theories the heat kernel expansion may be also
used to find classical action through the so-called spectral
action principle \cite{Chamseddine:1996zu}.

In the present paper we consider the case when the noncommutativity is
defined through the Moyal star product. If the operator $D$ contains
left (or right) Moyal multiplications only, the corresponding heat
kernel expansion was constructed in \cite{Vassilevich:2003yz}
on the torus, and in \cite{Gayral:2004ww} on the plane. Curved 
space\footnote{From other heat kernel calculations on curved
noncommutative spaces if the noncommutativity is realized in a
different way we would like to mention  the works
\cite{Avramidi:2004fc,Sasakura:2004dq}.}
modifications, conformal anomalies, and the Polyakov action were
calculated in \cite{Vassilevich:2004ym}. In the paper 
\cite{Gayral:2004cu} it was noted that if $D$ contains both
right and left Moyal multiplications (and if the matrix $\theta^{\mu\nu}$,
which defines the noncommutativity, is degenerate) the heat kernel
expansion contains contributions which are not local even in
the generalised noncommutative sense, so that some problems with
renormalizability can arise. Since the presence of both left and right
Moyal multiplications is required by many physical applications (to the
noncommutative $\phi^4$ theory, for example), we also consider such
operators, but throughout the paper we assume that $\theta^{\mu\nu}$
is non-degenerate\footnote{The fact that non-degenracy
of $\theta$ improves renormalisation was noted in \cite{Gayral:2004cu}.
This analysis was then extended to a position-dependent $\theta^{\mu\nu}$
in \cite{Gayral:2005ih}.}. In the next section we propose a method 
for calculation
of the heat kernel coefficients, and present explicit expressions for
several leading terms. We find that there are two types of the coefficients.
Leading coefficients are star-local (i.e., they are integrals of
star-polynomials of the fields), but there are also rather non-standard
non-local contributions, which are similar to the non-planar diagrams.
We the help of the heat kernel expansion we construct the large mass
expansion of the effective action and of the vacuum energy 
(sec.\ \ref{sec-ef}). This
expansion appears to be a strong noncommutativity expansion (i.e.,
it is valid for large $\theta$). Axial anomalies are considered in
sec.\ \ref{sec-an}, where we also find an anomaly-free model.
\section{Heat kernel}\label{sec-hk}
The Moyal product of two functions $f$ and $g$ on $\mathbb{R}^n$ can
be defined by the equation
\begin{equation}
f\star g = f(x) \exp \left( \frac i2 \, \theta^{\mu\nu}
\overleftarrow{\partial}_\mu \overrightarrow{\partial}_\nu \right)
g(x) \,.\label{starprod}
\end{equation}
$\theta$ is a constant antisymmetric matrix.
In this form the star product
has to be applied to plane waves and then extended
to all (square integrable) functions by means of the Fourier series.
The following properties of the Moyal star product will be used in
this paper
\begin{equation}
\int d^nx\, f\star g=\int d^nx\, f\cdot g\,,\qquad
\int d^nx\, f\star g\star h=\int d^nx\, h\star f\star g\,.
\label{closed}
\end{equation}

In this paper we consider the operators which can be represented
in the form
\begin{equation}
D=-(\nabla_\mu^2 + E) \,,\label{operD}
\end{equation}
where
\begin{eqnarray}
&&\nabla_\mu = \partial_\mu + L(\lambda_\mu )+R(\rho_\mu)\,,\label{covdir}\\
&&E=L(l_1)+R(r_1)+L(l_2) \circ R(r_2) \,.\label{defE}
\end{eqnarray}
$L$ and $R$ are operators of left and right Moyal multiplications,
\begin{equation}
L(l)\,f=l\star f,\qquad R(r)\,f =f\star r \,.\label{defLR}
\end{equation}
The operator $D$ we consider in this section acts on scalars.
Additional matrix structure does not lead to much complications
(see sec.\ \ref{sec-an}).  We use the symbol $\circ$ for the
operator products to distinguish from the star product of functions.
We shall, however, omit $\circ$ at some places
if this cannot lead to a confusion. 
We suppose that all fields $\lambda_\mu$, $\rho_\mu$, $l_{1,2}$,
$r_{1,2}$ fall off at the infinity faster than any power of a
radial coordinate on $\mathbb{R}^n$. The mass dimension of 
$\lambda_\mu$, $\rho_\mu$, $l_2$ and $r_2$ is one. The fields
$l_1$ and $r_1$ have mass dimension two.

Consider an $L_2$ scalar product on the space of functions
\begin{equation}
(f,g)=\int d^nx\, f^*(x) g(x) \,,\label{scaprod}
\end{equation}
where $f^*$ is a complex conjugate of $f$. With respect to this
product formal adjoints of the multiplication operators read
$R(r)^\dag =R(r^*)$, $L(l)^\dag = L(l^*)$. 

There are two independent gauge symmetries of this problem (cf. 
\cite{Liao:2004mu} where this ``double gauging'' phenomenon was
discovered in a different context):
\begin{equation}
D\to L(U_L^{-1})\circ D \circ L(U_L)\,, \quad {\mbox{and}}\quad 
D\to R(U_R^{-1})\circ D \circ R(U_R)\,.\label{gausim}
\end{equation}
The fields in the operator $D$ transform according to the following
rules
\begin{eqnarray}
&&\lambda_\mu \to U_L^{-1}\star \partial_\mu U_L 
+ U_L^{-1}\star \lambda_\mu \star U_L\,,
\nonumber\\
&&l_{1,2}\to U_L^{-1}\star l_{1,2} \star U_L\,,\nonumber\\
&&\rho_\mu \to \partial_\mu U_R \star U_R^{-1}+
U_R\star \rho_\mu \star U_R^{-1}\,,\nonumber\\
&&r_{1,2}\to 
U_R\star r_{1,2} \star U_R^{-1}\,.\label{fieldtrafo}
\end{eqnarray}
The left fields $\lambda_\mu$, $l_1$, $l_2$ (respectively, the right
fields $\rho_\mu$, $r_1$, $r_2$) are invariant under the transformations
parametrised by $U_R$ (respectively, by $U_L$). By representing 
$U_{L,R}=e^{w_{L,R}}$ and restricting (\ref{fieldtrafo}) to the
linear order in $w_{L,R}$ one obtains an infinitesimal version of
the gauge transformations
\begin{eqnarray}
&&\delta \lambda_\mu =  \partial_\mu w_L +[\lambda_\mu,w_L]\,,
\nonumber\\
&&\delta l_{1,2}=[l_{1,2},w_L],\nonumber\\
&&\delta \rho_\mu = \partial_\mu w_R + [w_R,\rho_\mu],\nonumber\\
&&\delta r_{1,2}=[w_R,r_{1,2}]\,.\label{infigau}
\end{eqnarray}
In the exponential and in the commutators all products are the Moyal
star products.

The gauge fields, $\lambda_\mu$ and $\rho_\mu$, and the gauge 
parameters $w_{L,R}$ are typically pure imaginary, so that
$U_{L,R}^{-1}=U_{L,R}^*$. There are two different field strengths
\begin{equation}
\nabla_\mu \circ \nabla_\nu - \nabla_\nu \circ \nabla_\mu
=L(\Omega_{\mu\nu}^L)+R(\Omega_{\mu\nu}^R),\label{cocode}
\end{equation}
where
\begin{eqnarray}
&&\Omega_{\mu\nu}^L=\partial_\mu \lambda_\nu - \partial_\nu \lambda_\mu
+ [\lambda_\mu,\lambda_\nu]\,\nonumber\\
&&\Omega_{\mu\nu}^R=\partial_\mu \rho_\nu -\partial_\nu \rho_\mu 
+ [\rho_\nu,\rho_\mu] \,.\label{Omn}
\end{eqnarray}
We also introduce two covariant derivatives
\begin{equation}
\nabla_\mu^L=\partial_\mu + [\lambda_\mu,\ \cdot\ ],\qquad
\nabla_\mu^R=\partial_\mu + [\ \cdot\ ,\rho_\mu ]\,.\label{twoder}
\end{equation}

Let us now evaluate the asymptotic expansion of the trace of
heat kernel
\begin{equation}
K(t,D)={\mathrm{Tr}}_{L^2}\left( e^{-tD}-e^{-tD_0}\right) \,.\label{subhk}
\end{equation}
In order to remove a trivial volume divergence of the trace we have
subtracted the heat kernel of the ``free'' operator 
$D_0=-\partial_\mu^2$. Alternatively, this volume divergence can be
regularised by introducing a smearing function under the trace
(see sec.\ \ref{sec-an}). From now on we suppose that the matrix
$\theta^{\mu\nu}$ is non-degenerate. Consequently, the dimension $n$
must be even.

To evaluate the $L_2$ trace we sandwich the exponentials between
plane waves $e^{ikx}$, integrate over $x$ (to produce diagonal
matrix elements of the heat operator), and finally integrate over
$k$ (to calculate the trace).  
\begin{equation}
K(t,D)=\int d^nx\int \frac {d^nk}{(2\pi)^n} e^{-ikx}
\left( e^{-tD}-e^{-tD_0}\right) e^{ikx}\,. \label{ktr}
\end{equation}
In order to evaluate the asymptotic expansion of (\ref{ktr})
at $t\to +0$ one has to extract the factor $e^{-tk^2}$.
\begin{eqnarray}
K(t,D)&=&\int d^nx\int \frac {d^nk}{(2\pi)^n} e^{-tk^2}\nonumber \\
&&\times \langle \exp \left( t\left( (\nabla -ik)^2 
+2i k^\mu (\nabla_\mu -ik_\mu)
+E\right)\right)-1\rangle_k ,\label{expk2}
\end{eqnarray} 
where we defined
\begin{equation}
\langle F \rangle_k \equiv e^{-ikx}\star F e^{ikx} \label{angle}
\end{equation}
for any operator $F$. Next one has to expand the exponential
in (\ref{expk2}) in a power series in $E$ and $(\nabla -ik)$.
Only a finite number of terms in this expansion contribute to
any finite order of $t$ in the $t\to +0$ asymptotic expansion of
the heat kernel. 

To illustrate the method let us consider the terms which are obtained
by expanding the exponential in (\ref{expk2}) up to the quadratic order in
$E$.
\begin{equation}
K(t,D)_{E^2}=\int d^nx\int \frac {d^nk}{(2\pi)^n} e^{-tk^2}
\langle tE + \frac {t^2}2 E^2 \rangle_k \,.\label{E2hk}
\end{equation}
Equation (\ref{defE}) yields
\begin{eqnarray}
&&E^2=R(r_1\star r_1)+L(l_1\star l_1) +R(r_2\star r_2)\circ L(l_2\star l_2)
+2R(r_1)\circ L(l_1)\nonumber\\
&&\qquad +R(\{ r_1,r_2\})\circ L(l_2)+R(r_2)\circ L(\{l_1,l_2\}) \,.\label{Esq}
\end{eqnarray}
The terms containing left or right multiplications only can be dealt with
easily by using the identities 
\begin{equation}
\int d^nx \langle R(r) \rangle_k = \int d^nx\, r(x)\,,\qquad
\int d^nx \langle L(l) \rangle_k = \int d^nx\, l(x)\label{Rrr}
\end{equation}
and the integral
\begin{equation}
\int \frac {d^nk}{(2\pi)^n} e^{-tk^2}=(4\pi t)^{-n/2}\,.\label{intk}
\end{equation}

The terms containing both left and right multiplications are somewhat
more difficult.
Consider a typical contribution of that type 
\begin{equation}
T(l,r) =\int d^nx\int \frac {d^nk}{(2\pi)^n} e^{-tk^2} \langle
L(l)\circ R(r) \rangle_k \label{ex1} 
\end{equation}
with some functions $r(x)$ and $l(x)$.
Let us expand $r(x)$ and $l(x)$ in the Fourier integrals
\begin{eqnarray}
&&r(x) = \frac 1{(2\pi)^{n/2}} \int d^nq\, r(q) e^{iqx},\nonumber\\
&&l(x)= \frac 1{(2\pi)^{n/2}} \int d^nq'\, l(q') e^{iq'x} .
\label{Fourier} 
\end{eqnarray}
Then
\begin{equation}
\langle L(l)\circ R(r) \rangle_k =\frac 1{(2\pi)^n} \int d^nq\, d^nq'\,
r(q)l(q') e^{i(q+q')x} e^{\frac i2 k\wedge (q-q')}
e^{-\frac i2 (q'-k)\wedge (q+k)} \,,\label{fourLR}
\end{equation}
where
\begin{equation}
k\wedge q\equiv \theta^{\mu\nu} k_\mu q_\nu \,.\label{dwedge}
\end{equation}
Next we substitute (\ref{fourLR}) in (\ref{ex1}) 
and integrate over $x$ and $q'$ to obtain
\begin{equation}
T(l,r)=  \int \frac {d^nk\, d^nq}{(2\pi)^n} e^{-tk^2}
l(-q)\, r(q) e^{-ik\wedge q}\,.\label{ex2}
\end{equation}
To calculate the integral over $k$ we complete the square in the exponential
\begin{equation}
-tk^2-ik\wedge q= -t\left( k_\mu +\frac i{2t} \theta_\mu^{\ \, \nu}q_\nu 
\right)^2 -\frac 1{4t} \theta^{\mu\mu'}\theta^\nu_{\ \, \mu'} q_\mu q_\nu \,.
\label{square}
\end{equation}
Then equation (\ref{ex2}) becomes
\begin{equation}
T(l,r) =\int \frac {d^n q}{(4\pi t)^{n/2}} l(-q) r(q)
\exp \left( -\frac 1{4t} \theta^{\mu\mu'}\theta^\nu_{\ \, \mu'} q_\mu q_\nu 
\right)\,.\label{ex3}
\end{equation}
To evaluate the asymptotic behaviour of (\ref{ex3}) at $t\to +0$ one has
to expand $l(-q) r(q)$ in Taylor series and then perform the Gaussian
integral over $q$. First two terms of this expansion read
\begin{equation}
T(l,r)=\left( \det \theta \right)^{-1} \left[ l(0) r(0)
+ t \left (\theta \theta^T \right)^{-1}_{\mu\nu}
\frac {\partial^2}{\partial q_\mu \partial q_\nu} 
\left( l(-q)r(q) \right)_{q=0} + \mathcal{O}(t^2) \right] \,.\label{ex4}
\end{equation}
Next one returns to the coordinate representation by using the following
formulae
\begin{eqnarray}
&&l(0) r(0)= \frac 1{(2\pi)^n} \int d^nx\, l(x)\,\int d^ny\, r(y)\,,
\label{coord1} \\
&&\frac {\partial^2}{\partial q_\mu \partial q_\nu} 
\left( l(-q)r(q) \right)_{q=0}= \frac 1{(2\pi)^n}
\left[ -\int d^nx\,x^\mu x^\nu l(x)\,\int d^ny\, r(y)\right.
\nonumber \\
&&\qquad\qquad 
 -\int d^nx\, l(x)\,\int d^ny\,y^\mu y^\nu r(y) 
+\int d^nx\,x^\mu l(x)\,\int d^ny\, y^\nu r(y) \nonumber\\
&&\qquad\qquad \left. 
+\int d^nx\,x^\nu l(x)\,\int d^ny\, y^\mu r(y)
\right].\label{coord2}
\end{eqnarray}
Note that both expressions in (\ref{coord1}) and
(\ref{coord2}) are invariant under constant
shifts of the coordinates $x$ and $y$, i.e. they do not depend
on the point where we put an origin of our coordinate system.
The integrals  in (\ref{coord2}) are not gauge invariant. We shall
return to this issue later.

By simply collecting the formulae given above one can easily find
the contributions of $E$ the heat kernel coefficients up to the
$E^2$ order (cf. (\ref{E2hk})). All other terms can be treated in
a similar way.  
In general case one can prove the following statements.
\begin{itemize}
\item[(i)] There is a power law asymptotic expansion of the heat kernel
(\ref{asymptotex}). All odd-numbered coefficients $a_k$ with $k=2j+1$
vanish, while the even-numbered coefficients with $k=2j$ are given
by the formula:
\begin{equation}
a_k(D)=a_k^L (D) + a_k^R(D) + a_k^{\rm mix}(D).\label{aaaa}
\end{equation}
\item[(ii)] The coefficients $a_k^L (D)$ (respectively, $a_k^R(D)$)
depend on the left (respectively, right) fields only. They are
represented by integral over $\mathbb{R}^n$ of gauge-invariant
star polynomials. These coefficients have been calculated earlier 
on \cite{Vassilevich:2003yz} (on the noncommutative torus) and 
in \cite{Gayral:2004ww} (on the noncommutative plane). First several
coefficients read
\begin{eqnarray}
&&a^L_2= (4\pi)^{-n/2} \int d^n x l_1(x) \,,\label{a2} \\
&&a^L_4= (4\pi)^{-n/2} \frac 1{12} \int d^n x  \left( 6
l_1 \star l_1 + \Omega^{L\mu\nu} \star \Omega^L_{\mu\nu} \right)
\label{a4}\\
&&a^L_6= (4\pi)^{-n/2} \frac 1{360} \int d^n x \left(
60 l_1\star l_1 \star l_1 + 30 l_1 \star \nabla^{L\mu}\nabla^L_\mu
 l_1  \right. \nonumber\\
&&\qquad\qquad\qquad 
+30 l_1 \star \Omega^L_{\mu\nu} \star \Omega^{L\mu\nu} 
- 4(\nabla^L_\sigma \Omega^L_{\mu\nu}) \star (\nabla^{L\sigma}\Omega^{L\mu\nu})
\label{a6}\\ 
&&\qquad\qquad\qquad \left.
+ 2 (\nabla^{L\nu}\Omega^L_{\mu\nu}) \star (\nabla^L_\sigma
\Omega^{L\mu\sigma})
-12 \Omega^L_{\mu\nu}\star \Omega^{L\nu}_{\ \ \,\sigma} \star 
\Omega^{L\sigma\mu}
\right),
\nonumber
\end{eqnarray}
The coefficients $a_k^R$ are obtained from $a_k^L$ by replacing
$\{l_1,\nabla^L,\Omega^L\}$ with $\{ r_1,\nabla^R,\Omega^R\}$
and the reversing the order of all multiplier under the integral\footnote{
This inversion rule follows from the identity 
$R(r)\circ R(r')=R(r'\star r)$. The inversion is partially taken care of
by the definitions of the covariant derivatives (\ref{twoder}) and
the field strengths (\ref{Omn}).}.
\item[(iii)] The mixed coefficients $a_k^{\rm mix}(D)$
vanish for $k\le n$. The first non-zero coefficient is
\begin{equation}
a_{n+2}^{\rm mix}=\frac{(\det \theta)^{-1}}{(2\pi)^n}
\left( \int d^nx\, l_2(x)\, \int d^ny\, r_2(y) +
2\int d^nx\, \lambda^\mu (x)\, \int d^ny\, \rho_\mu (y) \right) .\label{an2}
\end{equation}
Let $\lambda_\mu = \rho_\mu =0$. Then the next coefficient reads
\begin{eqnarray}
&&a_{n+4}^{\rm mix}=\frac{(\det \theta)^{-1}}{(2\pi)^n}
\left[ \frac 12 \int d^nx\, l_2^2\, \int d^ny\, r_2^2 +
\int d^nx\, l_1 \,\int d^ny\, r_1 \right.\nonumber\\
&&\qquad\qquad + \int d^nx\, l_2 \int d^ny\, r_1\star r_2 +
\int d^nx\, l_1\star l_2 \int d^ny\, r_2 \nonumber\\
&&\qquad\qquad + (\theta \theta^T)^{-1}_{\mu\nu} \left( 
- \int d^nx\, x^\mu \star x^\nu \star l_2 \,\int d^ny \, r_2
\right.\label{an4}\\
&&\qquad\qquad \left.\left. 
- \int d^nx\, l_2 \, \int d^ny \,y^\mu \star y^\nu \star r_2
+2 \int d^nx\, x^\mu l_2 \, \int d^ny \, y^\nu \star r_2
\right)\right] \nonumber
\end{eqnarray}
\end{itemize}

The coefficients (\ref{a2}) - (\ref{an2}) are gauge invariant.
The expression (\ref{an4}) is not gauge invariant since we 
assumed $\lambda_\mu =\rho_\mu =0$. If the gauge fields are
non-zero, the coordinates $x^{\mu}$ and $y^\mu$ which appear
under the integrals are replaced by the 
expressions
\begin{equation}
X^\mu = x^\mu + i\theta^{\mu\sigma} \lambda_\sigma \quad
\mbox{and} \quad Y^\mu =y^\mu -i\theta^{\mu\sigma} \rho_\sigma \label{XY}
\end{equation}
respectively. These shifted coordinates are gauge covariant,
$\delta X^\mu = [X^\mu ,w_L]$, $\delta Y^\mu = [w_R, Y^\mu]$.
Therefore, the gauge invariance of the heat kernel expansion 
is restored.

Note, that if $\theta^{\mu\nu}$ is degenerate, mixed contributions
to the heat kernel coefficients may appear earlier than in $a_{n+2}$,
which may affect renormalization \cite{Gayral:2004cu}. In the quantum
field theory language, mixed coefficients correspond to non-planar
diagrams. We shall discuss these points below.
\section{Effective action and vacuum energy}\label{sec-ef}
Let us first recall some basic facts regarding the one-loop effective
action in quantum field theory. Consider a scalar field $\phi$ described
by a classical action $S(\phi)$. In the background field formalism
one splits $\phi =\varphi + \delta\phi$, where $\varphi$ is a background
field. The field $\delta\phi$ describes quantum fluctuations. Then one
expands $S(\phi)$ about the background. The first term, $S(\varphi)$,
simply gives the classical approximation to the effective action,
The second term, which is proportional to the first derivative of 
$S(\varphi)$ is cancelled by external sources. The quadratic term
can be rewritten as
\begin{equation}
S^{(2)}=\int d^nx\, (\delta\phi) (D+m^2) (\delta \phi) \,,\label{S2}
\end{equation}
where $D$ is an operator which depends on the background field $\varphi$.
We have separated explicitly the mass term $m^2$. The functional integration
of $\exp (-S^{(2)})$ defines the one loop effective action
\begin{equation}
W=-\ln \int \mathcal{D}(\delta\phi) e^{-S^{(2)}}=
\frac 12 \ln \det (D+m^2)\,.\label{defW}
\end{equation} 
This expression is, of course, divergent and has to be regularised.
We use the zeta function regularization \cite{zetareg,Elizalde:1994gf}.
The zeta function is defined by the equation
\begin{equation}
\zeta(s,D+m^2)={\mathrm{Tr}}_{L^2} \left( (D+m^2)^{-s} - (D_0+m^2)^{-s}
\right),\label{defz}
\end{equation}
where $s$ is a complex spectral parameter. Again, as in (\ref{subhk}),
we subtracted the zeta function of the free operator $D_0$ in order
to remove a trivial volume divergence.

The regularised effective action reads
\begin{equation}
W_s=-\frac 12 \tilde\mu^{2s} \Gamma (s) \zeta (s,D+m^2) \,.\label{Ws}
\end{equation}
The regularization is removed in the limit $s\to 0$. $\tilde\mu$
is a constant of the dimension of mass introduced to keep proper
dimension of the effective action. The heat kernel and the zeta
function are related by a Mellin transformation. One can rewrite
(\ref{Ws}) as
\begin{equation}
W_s=-\frac 12 \tilde\mu^{2s} \int_0^\infty \frac {dt}{t^{1-s}}
K(t,D)e^{-tm^2} \,.\label{Wst}
\end{equation}

If $m^2$ is large enough the integral over $t$ is convergent at the
upper limit. There are, however, divergences at the lower limit which
are defined by the heat kernel expansion. By substituting 
(\ref{asymptotex}) in (\ref{Wst}) one obtains
\begin{equation}
W_s\simeq -\frac 12 \left( \frac{\tilde\mu}m \right)^{2s} \sum_k 
\Gamma \left( s+ \frac{k-n}2 \right) m^{n-k} a_k(D) \,.\label{Wshk}
\end{equation}
Let us recall, that the coefficient $a_0(D)$ vanishes because of the
subtraction of the volume term, the odd-numbered coefficients
$a_{2j+1}$ vanish on manifolds without boundaries in both commutative
and non-commutative cases. Consequently, the summation in (\ref{Wshk})
runs over even positive integers, $k=2,4,\dots$ We have assumed that
$n$ is even. The gamma functions in (\ref{Wshk}) have poles at $s=0$
for $k\le n$, and the corresponding terms in the sum define one-loop
divergences. According to the results obtained in the previous section
the coefficients $a_k(D)$ with $k\le n$ are integrals of star-polynomials
constructed from fields and their derivatives, i.e. the divergences
have a structure typical to the classical actions. One may hope
therefore, that the divergences may be absorbed in a redefinition
of the coupling constants. Particular normalisation conditions
depend, of course, on the model in question. Since the divergent
terms are always proportional to non-negative powers of the mass,
the scheme based on the subtraction of leading terms in the 
$m\to\infty$ asymptotics \cite{Bordag:1998vs} should work anyway,
although its physical meaning is not always clear.

With the help of (\ref{Wshk}) one can also evaluate the large mass
expansion of the effective action. First several terms in (\ref{Wshk})
are divergent. Therefore, as discussed above, the corresponding terms
in the large mass expansion are defined by the renormalization
of the classical action on a given background. The terms with $k>n$
are non-divergent. They represent ``genuine'' quantum corrections
to the effective action (since corresponding structures may be
absent in the classical action). In these terms we can put $s=0$
thus obtaining the following expression
\begin{equation}
W^{[1/m]}=-\frac 12 \sum_{p=1}^\infty m^{-2p} (p-1)!\, a_{2p+n}(D).
\label{1m-gen}
\end{equation}
In the commutative case, all terms of these expansion are local
since the heat kernel coefficients are local. This property is clearly
lost in the noncommutative case.

As an example, let us
consider a real scalar field $\phi$ in four dimensions with the classical
action\footnote{Sometimes the $\phi^4$ action on noncommutative plane is
modified by an external oscillator potential \cite{Grosse:2003nw}.}
\begin{equation}
S=\int d^4x \left( \frac 12 (\partial_\mu \phi)^2 +
\frac 12 m^2 \phi^2 + \frac g{24} \phi\star\phi\star\phi\star\phi \right),
\label{p4act}
\end{equation}
where $g$ is a coupling constant. By expanding this action around a
background field $\varphi$ and keeping the terms which are
quadratic in the fluctuations $\delta\phi$ only one arrives
at (\ref{S2}) with (cf. \cite{Gayral:2004cu})
\begin{equation}
D=-\partial_\mu^2 + \frac g6 \left[ R(\varphi \star \varphi )+
L(\varphi \star \varphi )+L(\varphi )\circ R (\varphi )\right]\,.
\label{Dp4}
\end{equation}
This operator corresponds to the following choice in (\ref{operD}) -
(\ref{defE})
\begin{equation}
\lambda_\mu =\rho_\mu =0\,,\quad l_1=r_1=-\frac g6 \varphi \star \varphi
\,,\quad l_2=-r_2=\sqrt{\frac g6} \varphi \,.\label{choi}
\end{equation}

The terms with $k=2,4$ in (\ref{Wshk}) are divergent. Near $s=0$ they
read
\begin{eqnarray}
&&W^{\mathrm{div}}=-\frac 12 \left[ \left( -\frac 1s +\gamma_E -1
-\ln \left(\frac {\tilde\mu^2}{m^2} \right) \right) m^2 a_2 \right.
\nonumber\\
&&\qquad\quad \left. +\left( \frac 1s -\gamma_E +  
\ln \left(\frac {\tilde\mu^2}{m^2} \right)\right) a_4 \right] +\mathcal{O}(s),
\label{Wdiv}
\end{eqnarray}
where 
\begin{eqnarray}
&&a_2=-\frac g{24\pi^2} \int d^4x\, \varphi^2 \,,\nonumber\\
&&a_4=\frac 1{16\pi^2}\, \frac {g^2}{36} \int d^4x \,
\varphi\star\varphi\star\varphi\star\varphi \,.\label{a24p4}
\end{eqnarray}
We see now that the divergences in (\ref{Wdiv}) can indeed be absorbed
in redefinitions of the mass and of the coupling constant in the classical
action (\ref{p4act}) (which is very well known).

By substituting (\ref{choi}) in (\ref{a6}) - (\ref{an4}) and then in
(\ref{1m-gen}) one obtains
\begin{eqnarray}
&&W^{[1/m]}=-\frac 1{32\pi^2}\, \frac {g^2}{648m^2}
\int d^4x \left( -g\varphi_\star^6 +3\varphi_\star^2 \partial_\mu^2
\varphi_\star^2 \right)\nonumber \\
&&\qquad\quad +\frac{(\det \theta)^{-1}}{(2\pi)^4m^2}\left[ \frac g{12}
\left( \int d^4x\, \varphi \right)^2 -\frac{g^2}{48m^2}
\left( \int d^4x\, \varphi^2 \right)^2 \right.\nonumber\\
&&\qquad\quad -\frac{g^2}{36m^2}\int d^4 x\ \varphi
\int d^4 y\, \varphi_\star^3 -\frac g{6m^2} (\theta\theta^T)^{-1}_{\mu\nu}
\left( \int d^4 x\, x^\mu \star x^\nu \star \varphi \int d^4y\,
\varphi \right.\nonumber\\
&&\qquad\quad \left.\left. -\int d^4x\, x^\mu \star \varphi
\int d^4y \, y^\nu \star \varphi \right) \right]+\dots \label{1mn4}
\end{eqnarray}
where $\varphi_\star^k$ is the $k$th star-power of $\varphi$. 
E.g., $\varphi_\star^3\equiv \varphi\star\varphi\star\varphi$.

As in the commutative case, this expansion is valid if the fields
and their derivatives are small compare to the mass. In the noncommutative
case $\theta^{\mu\nu}$ appears in the denominator, so one should
also assume that $\theta^{-1}$ is small in a natural
scale defined by the mass. Therefore,
we have constructed a {\em strong coupling expansion} with respect to
the noncommutativity parameter. Consequently, there is no smooth transition
to the case of a degenerate $\theta^{\mu\nu}$.

With the same technical tools we can also evaluate the vacuum
(Casimir) energies\footnote{For a recent review on the Casimir
energy see \cite{Bordag:2001qi}. Here we follow Ref.\ \cite{Bordag:2004rx}.} 
of static $n+1$ dimensional systems.
We still assume that $n$ is even, that the noncommutativity is confined
to $n$ spatial dimensions, and that $\theta$ is a non-degenerate
$n\times n$ matrix.
In the zeta-function regularization the ground state energy is defined
as
\begin{equation}
\mathcal{E}_s =\frac 12 \mu^{2s} \sum_p \varepsilon_p^{1-2s},
\label{Es}
\end{equation}
where $\varepsilon_p$ are eigenfrequencies of elementary excitations
defined as square root of eigenvalues of the Hamiltonian
\begin{equation}
H=D+m^2.\label{HDm}
\end{equation}
Formally taking $s=0$ in (\ref{Es}) yields just a sum of zero point
energies of elementary oscillators. We rewrite Eq.\ (\ref{Es})
through the zeta function of $H$,
\begin{equation}
\mathcal{E}_s =\frac 12 \mu^{2s} \zeta \left( s-\frac 12 ,H\right).
\label{Eszeta}
\end{equation}
The zeta function can be expressed in terms of the heat kernel
\begin{equation}
 \zeta \left( s-\frac 12 ,H\right) 
=\frac 1{\Gamma \left( s-\frac 12 \right)} \int_0^\infty 
dt\, t^{s-3/2} K(t,D)e^{-tm^2}.\label{zehk}
\end{equation}

We assume that the operator $D$ is as in sec.\ \ref{sec-hk}, so
that the expansion (\ref{asymptotex}) with integer powers of
$t$ exists. Then (\ref{zehk}) is finite at $s=0$ and a large mass
expansion of the vacuum energy exists without any infinite
renormalization of the couplings\footnote{This property always
holds in commutative zeta-regularised theories in odd dimensions.
In the noncommutative case it is essential that $\theta$ is
non-degenerate. Note that in both commutative and noncommutative
cases a finite renormalization of couplings may occur.}
\begin{equation}
{\mathcal{E}}^{[1/m]}=-\frac 1{4\sqrt{\pi}} \sum_{p=1}^\infty
a_{2p}(D)\, m^{n+1-2p} \Gamma \left( p-\frac{n+1}2 \right)
\label{E1m}
\end{equation}

As an example we consider a model in $2+1$ dimensions ($n=2$).
In this case, $\theta^{ij}=\Theta \epsilon^{ij}$, where
$\theta$ is a constant and $\epsilon^{ij}$ is the Levi-Civita
tensor. $i,j=1,2$ are the space-like indices. Again we consider
the $\phi^4$ theory, so that the operator is given by Eq.\ 
(\ref{Dp4}) with $\mu$, $\nu$ replaced by the two-dimensional
indices $i$, $j$.
\begin{eqnarray}
&&{\mathcal{E}}^{[1/m]}=-\frac{gm}{24\pi} \int d^2x\, \varphi^2 
-\frac{g^2}{576\pi m}\int d^2x\, \varphi_\star^4 \nonumber\\
&&\qquad\qquad +\frac{g}{96\pi^2 m} \Theta^{-2} 
\left[ \int d^2x\, \varphi \right]^2 +\dots \label{Ep4}
\end{eqnarray}
This expansion is valid if $\varphi$, its derivatives, and $1/\Theta$
are small compared to the mass. It would be interesting to apply
this expansion to quantum corrections to noncommutative solitons.
In the commutative case the heat kernel methods give rapidly
convergent series for the mass shift even if there is no
explicitly small parameter in the model \cite{AlonsoIzquierdo:2004ru}.
For supersymmetric solitons the heat kernel methods can even give
exact results for the mass shift \cite{Bordag:2002dg}, but they 
require supersymmetric boundary conditions on quantum fluctuations.

The very appearance of mixed terms in the heat kernel expansion
and of the corresponding terms in the effective action is a consequence
of qualitatively different behaviour of planar and non-planar diagrams
in noncommutative theories and of the UV/IR (ultraviolet/infrared)
mixing phenomenon  \cite{Chepelev:1999tt,Minwalla:1999px,Aref'eva:1999sn}.
There are some similarities between general structure of the mixed
heat kernel coefficients and that of non-planar diagrams. The heat
kernel expansion provides a systematic (and relatively simple) 
way to construct a large $m$
and strong noncommutativity expansion of the one-loop effective
action\footnote{A diagrammatic approach to the effective action
in the noncommuative $\phi^4$ theory can be found in \cite{Micu:2000xj}.}.

It is essential here that the space has a trivial topology. For a
non-trivial topology many inetersting effects may occur 
\cite{Chaichian:2001pw}.
\section{Localised heat kernel and anomalies}\label{sec-an}
Let us consider a classical action for the Dirac spinors on the
Moyal plane
\begin{equation}
S_\psi =\int d^n x \, \bar\psi \star \Dir \psi \,,\label{Spsi}
\end{equation}
where, in the Euclidean space, $\bar\psi =\psi^\dag$. We choose the
Dirac operator in the form
\begin{equation}
\Dir = i\gamma^\mu \left( \partial_\mu + iL(V_\mu^L)+
iR(V_\mu^R) +\gamma_5 L(A_\mu^L)+\gamma_5 R(A_\mu^R) \right)\,.
\label{Dirac}
\end{equation}
The Dirac gamma matrices satisfy the Clifford relation
$\gamma^\mu\gamma^\nu +\gamma^\nu\gamma^\mu =2\delta^{\mu\nu}$.
Independently of the dimension $n$ the chirality matrix is denoted
by $\gamma_5$, and $\gamma^\mu\gamma_5=-\gamma_5\gamma^\mu$,
$\gamma_5^\dag =\gamma_5$, $\gamma_5^2=1$.

The classical action (\ref{Spsi}) is invariant under a two parameter
family of infinitesimal gauge transformations
\begin{eqnarray}
&&\delta_w \psi =iw_L \star \psi + i \psi \star w_R,\qquad
\delta_w \bar \psi =-i\bar\psi \star w_L -iw_R\star \bar\psi ,\nonumber\\
&&\delta_w V_\mu^L=-\partial_\mu w_L -i[V_\mu^L,w_L],\nonumber\\
&&\delta_w V_\mu^R=-\partial_\mu w_R +i[V_\mu^R,w_R],\label{psiga}\\
&&\delta_w A_\mu^L=-i[A_\mu^L,w_L],\qquad
\delta_w A_\mu^R=i[A_\mu^R,w_R]\nonumber
\end{eqnarray}
and under a two-parameter family of infinitesimal axial gauge transformations
\begin{eqnarray}
&&\delta_\sigma \psi = \sigma_L \star \gamma_5 \psi +
\gamma_5 \psi \star \sigma_R,\qquad
\delta_\sigma \bar\psi =\sigma_R \star \bar \psi \gamma_5 +
\bar \psi \gamma_5 \star \sigma_L, \nonumber \\
&&\delta_\sigma V_\mu^L=i[A_\mu^L,\sigma_L],\qquad
\delta_\sigma V_\mu^R=-i[A_\mu^R,\sigma_R],\nonumber\\
&&\delta_\sigma A_\mu^L=-\partial_\mu \sigma_L -i[V_\mu^L,\sigma_L],
\label{psiax}\\
&&\delta_\sigma A_\mu^R=-\partial_\mu \sigma_R +i[V_\mu^R,\sigma_R].
\nonumber
\end{eqnarray}
Here we assume that $V_\mu^{L,R}$, $A_\mu^{L,R}$, $w_{L,R}$ and $\sigma_{L,R}$
are real. This explains some extra factors of $i$ as compared to sec.\ 
\ref{sec-hk}.

Now we like to define which of the symmetries (\ref{psiga}) and
(\ref{psiax}) are anomalous and calculate corresponding anomalies.
After integrating out the fermions one arrives at the effective
action
\begin{equation}
W^{[f]}=-\ln \det \Dir = -\frac 12 \ln \det \Dir^2 \,.\label{Wfer}
\end{equation}
Again we use the zeta-function methods to regularise the determinant
(\ref{Wfer}). We write the regularised effective action as
(cf. (\ref{Ws}))
\begin{equation}
W^{[f]}_{s}=\frac 12 \tilde\mu^2 \Gamma (s) \zeta (s,\Dir^2). 
\label{zetaDir}\end{equation}
Up to a certain point the calculation of the anomaly in the zeta-function
regularization goes precisely the same way as in the commutative
case \cite{Balachandran:1981cs,Wipf,Marachevsky:2003zb}.
After subtracting the pole at $s=0$ the effective action becomes
\begin{equation}
W^{[f]}
=\frac 12 \zeta (0,\Dir^2)'+\frac 12 \ln (\mu^2) \zeta (0,\Dir^2)\,,
\label{Wsub}
\end{equation}
where prime denotes differentiation with respect to $s$.
The renormalization ambiguity resides now in the constant $\mu^2$,
which has to be determined through a normalisation condition.

The variation of the zeta function induced by any variation of
the external fields $V_\mu^{L,R}$, $A_\mu^{L,R}$ in $\Dir$ is 
\begin{equation}
\delta \zeta (s,\Dir^2)=-2s {\rm Tr}\left( (\delta \Dir) \Dir
 \Dir^{2(-s-1)} \right).\label{deze}
\end{equation}
The variation of the Dirac operator under the gauge transformations
(\ref{psiga}) reads
\begin{equation}
\delta_w\Dir = -i [\Dir ,L(w_L)+R(w_R)]\,.\label{Dirga}
\end{equation}
The substitution of (\ref{Dirga}) in (\ref{deze}) yields
$\delta_w \zeta (s,\Dir^2)=0$ (where we used cyclic symmetry of
the trace). Consequently, $\delta_w W^{[f]}=0$. We conclude
that the zeta-function regularization preserves gauge symmetries.

For the axial gauge transformations (\ref{psiax}) we have
\begin{equation}
\delta_\sigma \Dir = - \{ \Dir ,\gamma_5 (L(\sigma_L)+R(\sigma_R))\}
\label{Dirax}
\end{equation}
so that the variation of the zeta function reads
\begin{equation}
\delta_\sigma \zeta (s,\Dir^2) =4s {\rm Tr}\left(
\gamma_5 (L(\sigma_L)+R(\sigma_R)) \Dir^{-2s} \right) 
\equiv 4s \zeta \left( \gamma_5 (L(\sigma_L)+R(\sigma_R)), s, \Dir^2 \right),
\label{zeax}
\end{equation}
where we defined a smeared (or localised) zeta function
\begin{equation}
\zeta (F,s,D)\equiv {\rm Tr}\left( F D^{-s} \right).\label{smez}
\end{equation}
In a similar way also a smeared heat kernel can be defined
\begin{equation}
K(F,t,D)\equiv {\rm Tr}\left( F e^{-tD} \right).\label{smehk}
\end{equation}
We assume that $F$ is a zeroth order operator (i.e., it contains
multiplication, but no explicit partial derivatives). As we will see
below, there is an asymptotic expansion as in the unsmeared
case
\begin{equation}
K(F,t,D)\simeq \sum_{k=0}^\infty t^{(k-n)/2} a_k (F,D).\label{smasym}
\end{equation}
 There is no need to subtract from the smeared heat kernel
the contribution of the ``free'' operator (cf. (\ref{subhk})) since
the volume divergences can be removed if the smearing functions
$\sigma_L$ and $\sigma_R$ fall off sufficiently fast at the infinity.
Moreover, in the present case ${\rm Tr} \left(\gamma_5 (L(\sigma_L)
+R(\sigma_R)) e^{-tD_0} \right)=0$ because of the presence of
$\gamma_5$ under the trace.

The heat kernel and the zeta function are related through the Mellin
transform. In particular, 
\begin{equation}
a_k(F,D)={\rm Ress}_{s=(n-k)/2} (\Gamma (s) \zeta (F,s,D))
\label{akRes}
\end{equation}
and $a_n=\zeta (F,0,D)$.

Finally one arrives at the following expression for the axial anomaly
\begin{equation}
{\mathcal{A}}_\sigma \equiv \delta_\sigma W^{[f]} = 
2 a_n \left( \gamma_5 (L(\sigma_L)+R(\sigma_R)). \Dir^2 \right)
\label{Aan}
\end{equation}
The Fujikawa approach \cite{Fujikawa:1979ay} and the
finite-mode regularization method \cite{Andrianov:1983fg}
give a similar expression for the anomaly (up to some peculiarities
arising from the presence of a dimensional regularization parameter
in that schemes).

The square of the Dirac operator can be represented in a form
similar to (\ref{operD}): 
\begin{equation}
\Dir^2=-((\partial_\mu +\omega_\mu)^2 + E)\,,\label{Dir2}
\end{equation}
where
\begin{eqnarray}
&&\omega_\mu = iL(V_\mu^L)+iR(V_\mu^R) +
\frac 12 [\gamma^\mu,\gamma^\nu]\gamma_5 \left( L(A_\nu^L)+R(A_\nu^R)
\right),\label{Dircon}\\
&&E=\frac i4 [\gamma^\mu,\gamma^\nu ] \left( L(V_{\mu\nu}^L)+
R(V_{\mu\nu}^R)\right)+\gamma_5 \left( L(\nabla^{\mu L}A_\mu^L)
+R(\nabla^{\mu R}A_\mu^R) \right)\nonumber \\
&&\qquad +(n-2) \left( L(A^{\mu L}\star A_\mu^L)+R(A^{\mu R}\star A_\mu^R)
+2L(A_\mu^L) \circ R(A^{\mu R}) \right)\nonumber\\
&&\qquad +\frac 14 (n-3) [\gamma^\mu ,\gamma^\nu] \left( 
L([A_\mu^L,A_\nu^L]) + R([A_\nu^R,A_\mu^R]) \right).\label{DirE}
\end{eqnarray}
We have defined
\begin{eqnarray}
&&V_{\mu\nu}^L=\partial_\mu V_\nu^L - \partial_\nu V_\mu^L
+i[V_\mu^L,V_\nu^L],\qquad 
V_{\mu\nu}^R=\partial_\mu V_\nu^R - \partial_\nu V_\mu^R
+i[V_\nu^R,V_\mu^R],\nonumber\\
&&\nabla_\mu^L A_\nu^L=\partial_\mu A_\nu^L +i[V_\mu^L,A_\nu^L],
\qquad 
\nabla_\mu^R A_\nu^R=\partial_\mu A_\nu^R -i[V_\mu^R,A_\nu^R].
\label{VDA}
\end{eqnarray}

Now we are ready to calculate the heat kernel coefficient in (\ref{Aan}).
There are two differences to the case considered in sec.\ \ref{sec-hk}:
both $\omega_\mu$ and $E$ are matrix-valued, and the heat kernel
is smeared with a zeroth order operator. An extension to a matrix valued
$D$ goes almost without an effort. All steps go through, but one has
to replace $\lambda_\mu$, $\rho_\mu$, $l_{1,3}$ and $r_{1,2}$
by the matrices which follow from (\ref{Dircon}) and (\ref{DirE}).
Because of the matrix structure, the terms with right multiplications
do not commute any more with the terms with left multiplications, but
this effect is important in ``mixed'' coefficients only. The smearing
operator will appear linearly in all expressions, and one should not
forget to take a trace over the spinor indices. 
For example, the $E$-terms in the heat kernel (cf. (\ref{E2hk} in the
unsmeared case) up to the quadratic order in $E$ read
\begin{equation}
K(F,t,D)_{E^2}=\int d^nx\int \frac {d^nk}{(2\pi)^n} e^{-tk^2}
{\rm tr} \left\langle F\left( 1+ tE + \frac {t^2}2 E^2 \right) 
\right\rangle_k 
\,,\label{E2sm}
\end{equation}
where $\rm tr$ is the $\gamma$-matrix trace. For $n=2$ and
$n=4$ one can calculate the anomaly by expanding the exponents, as we
have outlines in sec.\ \ref{sec-hk} above. This way is the easiest
one from the conceptual point of view, but it is also rather lengthy.
More experienced readers can choose a different way based
on functorial properties of the heat kernel (some useful tools
can be found in \cite{Branson:1997ze}). First one proves that the
coefficient $a_n(F,D)$ does not contain mixed contributions as in
the unsmeared case, then one classifies the invariants of proper
dimension which may appear in $a_n(F,D)$. The numerical coefficients
in front of these invariants are then defined by comparing to
the results for noncommutative heat kernel \cite{Vassilevich:2003yz}
and for the commutative axial anomaly \cite{Andrianov:1983fg}.
Both methods, of course, give identical results.

In two dimensions the anomaly reads
\begin{eqnarray}
&&{\mathcal A}_{\sigma}=-\frac i\pi \int d^2x \left[ \sigma_L \star
\left( \frac 12 \epsilon^{\mu\nu} (iV_{\mu\nu}^L - [A_\mu^L,A_\nu^L])
+i \nabla_\mu^L A^{L\mu} \right) \right.\nonumber\\
&&\qquad\quad \left. +\sigma_R \star
\left( \frac 12 \epsilon^{\mu\nu} (iV_{\mu\nu}^R + [A_\mu^R,A_\nu^R])
+i \nabla_\mu^R A^{R\mu} \right) \right]. \label{2danom}
\end{eqnarray}
This result is consistent with earlier calculations 
\cite{Moreno:2000kt,Blas:2005ng} performed 
in the models either without $V_\mu^R$ \cite{Moreno:2000kt}, or 
when $V_\mu^R$ and $V_\mu^L$ act on different field components
\cite{Blas:2005ng}.

In four dimensions, it is useful to split the anomaly in four 
contributions 
\begin{eqnarray}
&&{\mathcal A}_{\sigma}={\mathcal A}_{\sigma_L}^+ +{\mathcal A}_{\sigma_L}^-
+{\mathcal A}_{\sigma_R}^+ + {\mathcal A}_{\sigma_R}^- ,\label{ALR}\\
&&{\mathcal A}_{\sigma_L}^+= \frac{-i}{24\pi^2} \int d^4x \sigma_L \star
\left( -4 [\nabla_\mu^L V^{L\mu\nu},A_\nu^L] +2 [\nabla_\mu^L A_\nu^L,
V^{L\mu\nu}] \right.\nonumber\\
&&\qquad\quad +2 i \nabla_\mu^L \nabla^{L\mu} \nabla_\nu^L A^{L\nu}
+4 i \{ \{ \nabla_\mu^L A_\nu^L,A^{L\nu}\},A^{L\mu} \} \nonumber\\
&&\qquad\quad +2 i \{ \nabla_\mu^L A^{L\mu},A_\nu^L \star A^{L\nu}\}
+4 i A_\mu^L \star (\nabla_\nu^L A^{L\nu} )\star A^{L\mu} \nonumber\\
&&\qquad\quad \left. +i [[A_\mu^L,A_\nu^L],A^{L\mu\nu}] \right)
\label{ALp}\\
&&{\mathcal A}_{\sigma_L}^-= \frac{-i}{48\pi^2} \int d^4x \sigma_L \star
\epsilon^{\mu\nu\rho\sigma} \left( 3i V_{\mu\nu}^L \star V_{\rho\sigma}^L
-i A_{\mu\nu}^L \star A_{\rho\sigma}^L\right.\nonumber\\
&&\qquad\quad -2 (V_{\mu\nu}^L\star A_\rho^L \star A_\sigma^L +
A_\mu^L \star A_\nu^L \star V_{\rho\sigma}^L ) -8 A_\mu^L \star
V_{\nu\rho}^L \star A_\sigma^L \nonumber\\
&&\qquad\quad \left. +4i A_\mu^L \star A_\nu^L \star A_\rho^L 
\star A_\sigma^L \right),\label{ALm}
\end{eqnarray}
where $A_{\mu\nu}^L=\nabla_\mu^L A_\nu^L -\nabla_\nu^L A_\mu^L$.

Due to the identity (cf. \cite{Andrianov:1983fg} for the commutative
case)
\begin{eqnarray}
&&{\mathcal{A}}_{\sigma_L}^+=\frac 1{48\pi^2}\, \delta_{\sigma_L} \,
\int d^4 x \left( -V_{\mu\nu}^L \star V^{L\mu\nu} 
-2 (\nabla_\mu^L A^{L\mu})\star (\nabla_\nu^L A^{L\nu})\right. \label{locct}\\
&&\qquad\quad \left. -4i A_\mu^L \star A_\nu^L \star V^{L\mu\nu}
-2 A_\mu^L \star A_\nu^L \star A^{L\mu}\star A^{L\nu}
+6 A_\mu^L \star A^{L\mu} \star A_\nu^L \star A^{L\nu} \right)\nonumber
\end{eqnarray}
the part ${\mathcal{A}}_{\sigma_L}^+$ can be cancelled by a star-local
counterterm. The other part, ${\mathcal{A}}_{\sigma_L}^-$, cannot be cancelled
by a counterterm. This part is called the topological part of the
anomaly. It is scheme-independent, and plays a more important role than
${\mathcal{A}}_{\sigma_L}^+$. Below we discuss the topological part
only.

The anomaly related to the $\sigma_R$ transformations can be obtained
by applying the same rule as in sec.\ \ref{sec-hk}: one has to take
(\ref{ALp}) and (\ref{ALm}), replace $L$ by $R$ everywhere, and invert the
order of all multipliers. The topological part of the anomaly reads 
\begin{eqnarray}
&&{\mathcal A}_{\sigma_R}^-= \frac{-i}{48\pi^2} \int d^4x \sigma_R \star
\epsilon^{\mu\nu\rho\sigma} \left( 3i V_{\mu\nu}^R \star V_{\rho\sigma}^R
-i A_{\mu\nu}^R \star A_{\rho\sigma}^R\right.\nonumber\\
&&\qquad\quad +2 (V_{\mu\nu}^R\star A_\rho^R \star A_\sigma^R +
A_\mu^R \star A_\nu^R \star V_{\rho\sigma}^R ) +8 A_\mu^R \star
V_{\nu\rho}^R \star A_\sigma^R \nonumber\\
&&\qquad\quad \left. +4i A_\mu^R \star A_\nu^R \star A_\rho^R 
\star A_\sigma^R \right).\label{ARm}
\end{eqnarray}

Again our results agree with previous calculations of the abelian anomalies
without $V^R$-fields and axial vector fields 
\cite{Ardalan:2000cy,Gracia-Bondia:2000pz,Banerjee:2001un,Armoni:2002fh}.
We also like to mention a couple of recent publications which
discuss the Fujikawa approach
\cite{Martin:2005jy} and the operator approach
\cite{Banerjee:2005zq} to the anomalies. Chiral anomaly on the noncommutative
torus was calculated in \cite{Iso:2002jc}.

If $\theta^{\mu\nu}$ is degenerate, the heat kernel coefficient
$a_n$ can contain mixed contributions. Consequently, non-planar
contributions to the anomaly appear \cite{Armoni:2002fh}.

We conclude this section by constructing
a model in four dimensions which has zero axial anomaly.
Let us consider the action (\ref{Spsi}) where we choose
\begin{equation}
A_\mu^L =-A_\mu^R\equiv A_\mu,\qquad V_\mu^L=-V_\mu^R \equiv V_\mu \,.
\label{asymf}
\end{equation}
Then also $V_{\mu\nu}^L=-V_{\mu\nu}^R$, $A_{\mu\nu}^L=-A_{\mu\nu}^R$, etc.
This model has a trivial commutative limit. The relations (\ref{asymf})
are preserved by gauge and axial gauge symmetries with the parameters
restricted according to the relation
\begin{equation}
w_L=-w_R,\qquad \sigma_L=-\sigma_R .\label{asympar}
\end{equation}
Obviously, the (topological part of the) anomaly is automatically
zero:
\begin{equation}
{\mathcal{A}}_{\sigma_L}^- + {\mathcal{A}}_{\sigma_R}^-=0.
\label{zeroan}
\end{equation} 

Cancellations of gauge anomalies in noncommutative theories were
discussed, e.g., in \cite{Intriligator:2001yu}.
\section{Conclusions}\label{sec-co}
In this paper we constructed the heat kernel expansion for the operators
which contain both left and right Moyal multiplications.
We found two types of the terms. The terms of the first type are
star-local and depend either on right or on left fields. The other
terms are non-local, and they contain mixtures of left and right 
fields. Next we applied our results to the $\phi^4$ theory and 
constructed a large mass and strong noncommutativity expansion of
the effective action and of the vacuum energy. Then we calculated the
axial anomaly, which do not contain mixed (non-planar) contributions
and, in fact, looks rather standard. We also found a model where the
topological part of the axial anomaly is identically zero.

Our work extends considerably the class of the operators on noncommutative
spaces for which the heat kernel expansion is known. Possible applications
of the results are not exhausted the the examples given above. It would be
interesting to consider the consequences for the spectral action principle
and to calculate quantum corrections to noncommutative instantons and solitons.
As a more formal development one can consider a degenerate 
noncommutativity parameter (cf. \cite{Gayral:2004cu}). In this case,
some non-planar contributions to the anomalies should appear leading
interesting physical consequences (cf. the discussion in
\cite{Ardalan:2000qk,Armoni:2002fh,Nakajima:2003an}).
\section*{Acknowledgements}
I am grateful to Peter Gilkey for helpful remarks.
This work was supported in part by the DFG project BO 1112/12-2.

\end{document}